\documentclass{UniDraft}

\usepackage[acronym]{glossaries}
\makeglossaries
\glsdisablehyper
\newacronym[plural=CPUs, longplural=central processing units]{cpu}{CPU}{central processing unit}
\newacronym{gpu}{GPU}{graphics processing unit}
\title{\bfseries \large Observation of Erratic Non-Hermitian Skin Localization and Transport}

\author[1,2]{Jia-Xin Zhong}
\author[2]{Jee Woo Kim}
\author[3,4,*]{Stefano Longhi}
\author[2,$\dagger$]{Yun Jing}

\affil[1]{Key Laboratory of Modern Acoustics and Institute of Acoustics, Nanjing University,
Nanjing 210093, China}
\affil[2]{Graduate Program in Acoustics, The Pennsylvania State University, University Park, PA 16802, USA}
\affil[3]{Dipartimento di Fisica, Politecnico di Milano, Piazza Leonardo da Vinci 32, I-20133 Milano, Italy}
\affil[4]{IFISC (UIB-CSIC), Instituto de Fisica Interdisciplinary Sistemas Complejos, E-07122 Palma de Mallorca, Spain}
\date{\vspace{-1em}\small 
    \textsuperscript{*}stefano.longhi@polimi.it; 
    \textsuperscript{$\dagger$}yqj5201@psu.edu
}

\titleformat{\section}         % 设置一级标题（Section）格式
  {\normalfont\Large\bfseries} % 字体格式：正常字体、大号、加粗
  {}                           % 【关键】标签留空，即不显示 "Section 1"
  {0pt}                        % 【关键】标签与标题文字之间的水平间距设为 0
  {}                           % 标题前的代码

\titleformat{\subsection}      % 设置二级标题（Subsection）格式
  {\normalfont\large\bfseries} % 字体格式：比一级小一点、加粗
  {}                           % 【关键】标签留空
  {0pt}                        % 【关键】间距设为 0
  {}
\titlespacing*{\section}{0pt}{3.5ex plus 1ex minus .2ex}{2.3ex plus .2ex}
\titlespacing*{\subsection}{0pt}{3.25ex plus 1ex minus .2ex}{1.5ex plus .2ex}
\usepackage{abstract}

\begin{document}
% Enable line numbers for review. Comment to disable.
% \linenumbers
% 把标题加入PDF书签，但不在目录里加上标题
\pdfbookmark[1]{Title}{title} % [1] 是书签层级，'title' 是内部引用标签

\maketitle
% \addcontentsline{toc}{part}{Title}
% enable Page 1 of xx at the first page

% \addcontentsline{toc}{section}{Title}

% \addcontentsline{toc}{section}{Contents}
% \vspace{-2em}

\begin{abstract}
Localization is a pervasive phenomenon across physics, shaping transport from electrons in solids to light and sound in engineered media.
% Well-known mechanisms such as Anderson localization impose predictable spatial confinement linked to disorder, while the non-Hermitian skin effect provides a striking example of boundary localization.
In traditional settings, disorder strongly impedes transport, resulting in dynamical localization or, at best, sub-ballistic or diffusive dynamics. 
A distinct and previously unobserved regime, erratic non-Hermitian skin localization (ENHSL), can arise in globally reciprocal non-Hermitian lattices with disorder. 
It features macroscopic, disorder-dependent localization at irregular bulk positions with subexponential decay, linked to stochastic interfaces governed by the universal order statistics of random walks. 
We realize this regime experimentally in an acoustic lattice implementing a disordered Hatano-Nelson chain with imaginary gauge fields.
Using Green's-function-based spectroscopy together with time-resolved measurements on the same platform, we reconstruct the full complex spectrum and eigenstates, and directly observe wave-packet dynamics.
Remarkably, we observe ballistic transport despite strong spectral localization. We develop a transport theory that connects the dominant propagation site to the maximal random-walk excursion within an expanding light cone and predicts a universal Lévy-arcsine statistics, in quantitative agreement with experiment. Our results decouple eigenstate localization from transport and establish ENHSL as a new paradigm for wave dynamics.
\end{abstract}

\thispagestyle{firststyle}

\newpage

\section{Introduction}
Wave localization is a unifying phenomenon across physics, shaping transport in systems ranging from electrons in solids to light, sound, and other engineered media. The most celebrated example, Anderson localization \cite{Anderson1958AbsenceDiffusionCertain, Thouless1974ElectronsDisorderedSystems,Evers2008AndersonTransitions} 
%\cite{A1,A2,A3}
, arises when disorder disrupts an underlying periodic structure, producing exponentially localized eigenstates and strongly inhibited motion. Even in special models where the localization length becomes unbounded, transport remains at most sub-ballistic \cite{delRio1995WhatLocalization, Dunlap1990AbsenceLocalizationRandomdimer, Wu1991PolyanilineRandomdimerModel}, and transitions to extended phases proceed via mobility edges through critical, fractal states \cite{Evers2008AndersonTransitions, Aoki1983CriticalBehaviourExtended}.

Non-Hermitian systems \cite{Ashida2020NonHermitianPhysics}%
%\cite{NH0}
—featuring gain, loss, or effective non-reciprocal couplings—have recently expanded the landscape of localization and transport phenomena. A paradigmatic example is the non-Hermitian skin effect (NHSE) in disorder-free lattices \cite{Yao2018EdgeStatesTopological, Kunst2018BiorthogonalBulkboundaryCorrespondence, Lee2019AnatomySkinModes,Okuma2020TopologicalOriginNonHermitian, Helbig2020GeneralizedBulkBoundary, Xiao2020NonHermitianBulkBoundary, Weidemann2020TopologicalFunnelingLight, Ghatak2020ObservationNonHermitianTopology, Wang2021DetectingNonBlochTopological, Bergholtz2021ExceptionalTopologyNonHermitian, Ding2022NonHermitianTopologyExceptionalpoint, Zhang2022ReviewNonHermitianSkin, Okuma2023NonHermitianTopologicalPhenomena, Lin2023TopologicalNonHermitianSkin}
where asymmetric couplings cause an extensive set of bulk modes to accumulate at system boundaries, overturning conventional bulk-boundary correspondence. In disordered non-Hermitian systems, Anderson localization also arises \cite{Hatano1996LocalizationTransitionsNonHermitian, Basiri2014LightLocalizationInduced, Tzortzakakis2020NonHermitianDisorderTwodimensional, Kawabata2021NonunitaryScalingTheory, Luo2021UniversalityClassesAnderson, Yusipov2017LocalizationOpenQuantum, Lin2022ObservationNonHermitianTopological, Weidemann2021CoexistenceDynamicalDelocalization} %
%\cite{NH1,NH2,NH3,NH4,NH5,NH6,NH7,NH8}; 
yet, unlike in Hermitian settings, wave propagation can persist even when all eigenstates remain exponentially localized with a bounded localization length. This counterintuitive transport is driven by the stochastic lifetimes of localized modes \cite{Weidemann2021CoexistenceDynamicalDelocalization, Tzortzakakis2021TransportSpectralFeatures, Leventis2022NonHermitianJumpsDisordered, Longhi2023AndersonLocalizationDissipative, Li2025UniversalNonHermitianTransport}
% \cite{NH8,NH9,NH10,NH11},
generating jump-like dynamics \cite{Weidemann2021CoexistenceDynamicalDelocalization, Leventis2022NonHermitianJumpsDisordered, Longhi2023AndersonLocalizationDissipative, Li2025UniversalNonHermitianTransport}
% \cite{NH8,NH10,NH11} 
and universal sub-ballistic spreading \cite{Longhi2023AndersonLocalizationDissipative}.
%\cite{NH11}.

Despite these differences, Anderson localization in both Hermitian and non-Hermitian systems is traditionally characterized by two shared hallmarks: exponentially localized eigenstates with spatial profiles dictated by disorder and stochastically centered throughout the system; and a corresponding suppression of ballistic transport, yielding either dynamical localization, diffusive or sub-ballistic motion. These features have long shaped our intuition about the interplay between disorder and wave propagation.

Here we experimentally demonstrate a fundamentally different localization regime—erratic {non-Hermitian} skin localization {(ENHSL)}—that emerges in globally reciprocal non-Hermitian lattices with disorder \cite{Longhi2025ErraticNonHermitianSkin}.
% Unlike Anderson localization, which yields exponentially localized modes with well-defined centers, or the NHSE, which forces modes to system boundaries, {ENHSL} produces macroscopic and highly irregular localization peaks whose positions vary unpredictably across disorder realizations.
Unlike Anderson localization or the NHSE, ENHSL produces macroscopic and highly irregular localization peaks whose positions vary unpredictably across disorder realizations.
The resulting spatial profiles display subexponential decay, are not fractal, and reflect stochastic interfaces governed by the order statistics of a symmetric random walk \cite{Longhi2025ErraticNonHermitianSkin}, establishing a direct link between non-Hermitian disorder and universal stochastic processes.

Realizing this phenomenon experimentally {is technically challenging because it} requires a platform with tunable non-Hermitian couplings, controllable disorder {over many realizations}, and global reciprocity, together with the {direct access to both spectral observables (complex eigenvalues and eigenstates) and time-resolved propagation.} 
Active acoustic lattices meet these requirements, offering tunable non-Hermitian couplings via pump-probe feedback, flexible disorder implementation, and high-accuracy wave field measurements \cite{Zhang2021AcousticNonHermitianSkin, Liu2022ExperimentalRealizationWeyl, Chen2025DirectMeasurementTopological, Tong2025ObservationFloquetBlochBraids, Zhong2025HigherorderSkinEffect, Zhong2025ExperimentallyProbingNonHermitian}.
Building on our Green's-function-based spectroscopy technique for reconstructing the full complex spectrum and eigenstates \cite{Zhong2025ExperimentallyProbingNonHermitian}, we further develop time-resolved measurements to track wave-packet dynamics, enabling a unified spectral-and-dynamical characterization of erratic localization and transport on the same platform.
Using this platform to implement a disordered Hatano–Nelson model with imaginary gauge fields, we directly observe the defining signatures of {ENHSL}: large-scale, sample-specific {bulk} localization peaks with strongly fluctuating positions across disorder realizations. 
Beyond establishing the spectral signatures of ENHSL, we develop and experimentally validate a complementary theory for wave-packet dynamics and transport that extends the original spectral picture of ENHSL.
Our measurements reveal that, despite strong spectral localization, the ensemble-averaged spreading remains ballistic.

Specifically, we find that wave excitations are drawn toward the site where the symmetric random-walk landscape attains its largest excursion within the light cone of the disorder-free lattice. This dominant site is a stochastic, time-dependent quantity whose distribution follows the universal Lévy arcsine law. This mechanism reveals that strong non-Hermitian localization does not necessarily impede ballistic motion—contradicting the long-standing assumption that disorder-induced confinement must inhibit transport or reduce it to sub-ballistic or diffusive behaviour.

\section{Results}

\subsection{Model and conceptual overview}

\begin{figure}[!htbp]
    \centering
    \phantomsection 
    \includegraphics[width=0.6\textwidth]{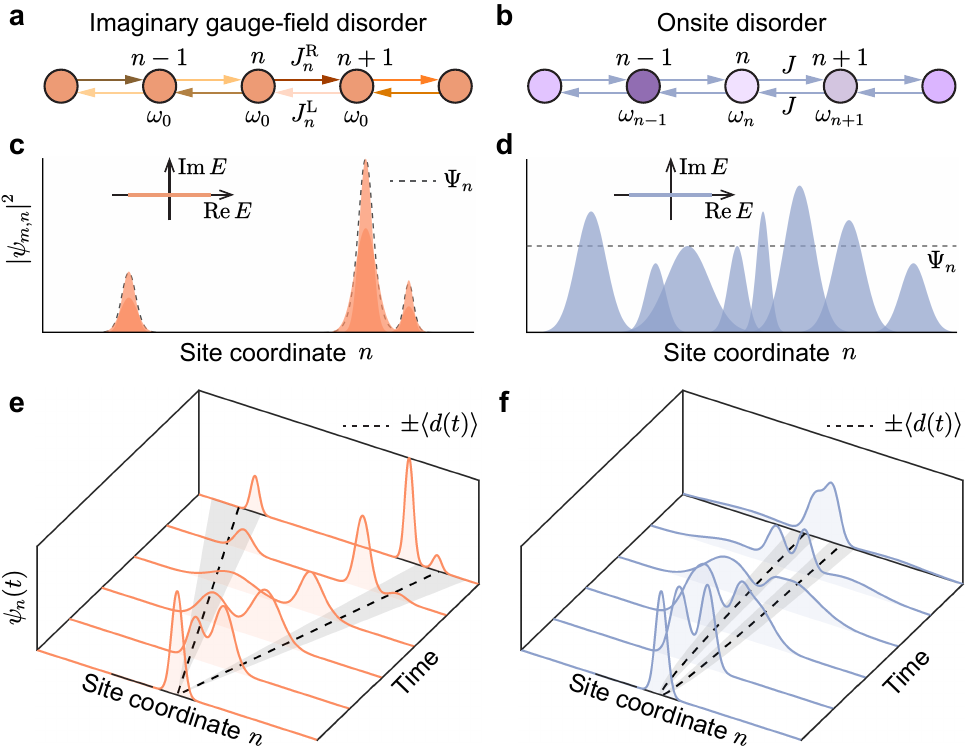}
    \caption{
        \textbf{Conceptual comparison between ENHSL and Hermitian Anderson localization.}
        \textbf{a}, Tight-binding schematic of a globally reciprocal non-Hermitian chain with disordered imaginary gauge fields (hopping disorder).
         Sites share a uniform onsite potential $(\omega_0)$, while the link colors indicate spatially fluctuating non-reciprocal hoppings ($J_n^{\mathrm{L/R}}$).
        \textbf{b}, Hermitian Anderson chain with onsite disorder (site colors, $\omega_n$) and Hermitian uniform hoppings ($J$).
        \textbf{c}, Spectral signatures of ENHSL. The inset sketches the complex spectrum, which remains (approximately) real despite non-reciprocal hoppings under global reciprocity. The main panel illustrates a typical ENHSL eigenstate profile featuring a dominant bulk peak accompanied by two weaker satellite peaks; the dashed curve indicates the summed eigenstate intensity $\Psi_n=\sum_m|\psi_{m,n}|^2$, which is dominated by the same peak structure. 
        \textbf{d}, Spectral signatures of Anderson localization: many exponentially localized eigenstates centered at different sites, yielding a relatively featureless mode-summed profile $\Psi_n$ (dashed line). 
        \textbf{e}, Dynamical response in ENHSL following a bulk-centered excitation. 
        The 3D traces show the spatiotemporal evolution of the wave-packet envelope $\psi_n(t)$ for a single disorder realization.
        Dashed guidelines show the symmetric ensemble-averaged spreading envelope, $n=\pm \langle d(t)\rangle$ (gray shading: realization-to-realization fluctuations), highlighting the ballistic transport.
        \textbf{f}, Dynamical response in the Anderson case, where the wave packet remains dynamically localized and the ensemble-averaged spreading saturates at long times (dashed guideline).
    }
    \label{fig:schematic}
    \addcontentsline{toc}{subsection}{Figure \thefigure}
\end{figure}

We consider a one-dimensional (1D) Hatano-Nelson chain \cite{Hatano1996LocalizationTransitionsNonHermitian} with spatially disordered imaginary gauge fields $h_n$ \cite{Longhi2025ErraticNonHermitianSkin,Midya2024TopologicalPhaseTransition}, sketched in Fig.~\ref{fig:schematic}a.
The defining feature is that local non-reciprocity fluctuates from link to link, yet the net imaginary gauge bias vanishes in each realization, i.e. $\sum_n h_n=0$ (global reciprocity).
In the tight-binding description, the right/left hoppings on link $n$ are
\begin{equation}
J^{\mathrm{R}}_{n} = J\,{e}^{h_n}, \qquad
J^{\mathrm{L}}_{n} = J\,{e}^{-h_n},
\label{eq:HN_hoppings}
\end{equation}
where $\{h_n\}$ are independent random variables drawn from a distribution $f(h)$ with zero mean. 
In this work, we sample $h_n$ from a uniform distribution $h_n\sim\mathcal{U}(-\Delta h,\Delta h)$, while the key dynamical predictions depend only on the symmetry and finite variance of $f(h)$.
A convenient way to represent the gauge disorder is via the cumulative gauge field (random-walk landscape)
\begin{equation}
X_n = \sum_{\ell=0}^{n-1} h_\ell,
\label{eq:random_walk_landscape}
\end{equation}
which maps the link disorder to a symmetric random walk in the site coordinate $n$.
According to the spectral theory of the ENHSL~\cite{Longhi2025ErraticNonHermitianSkin}, large excursions of $X_n$ give rise to stochastic interfaces that seed macroscopic localization structures through local skin localization. 
This leads to the natural question of whether the spectral localization inherent to the ENHSL permits any form of transport, or instead fully suppresses it. Although spectral localization is typically linked to dynamical localization or sub-ballistic transport \cite{delRio1995WhatLocalization, Dunlap1990AbsenceLocalizationRandomdimer, Wu1991PolyanilineRandomdimerModel,Weidemann2021CoexistenceDynamicalDelocalization, Tzortzakakis2021TransportSpectralFeatures, Leventis2022NonHermitianJumpsDisordered, Longhi2023AndersonLocalizationDissipative}, we show here that the extremal statistics of $X_n$ can, quite unexpectedly, enable on average ballistic transport.

To emphasize the genuinely new features of ENHSL, Fig.~\ref{fig:schematic} compares this globally reciprocal non-Hermitian model (Fig.~\ref{fig:schematic}a) with a conventional Hermitian (Anderson) disordered chain (Fig.~\ref{fig:schematic}b). %, where disorder enters as onsite energy fluctuations while hopping is uniform and reciprocal.
% We first summarize the spectral signatures associated with the two models.
In the globally reciprocal disordered Hatano-Nelson chain, ENHSL appears in each realization as macroscopic localization peaks at irregular \emph{bulk} positions, with all eigenstates displaying sub-exponential localization. A schematic eigenstate profile is shown in Fig.~\ref{fig:schematic}c: a dominant bulk peak is typically accompanied by a few weaker satellite peaks, producing a strongly nonuniform mode-summed intensity $\Psi_n=\sum_m |\psi_{m,n}|^2$, which is itself dominated by the same peak structure.
Here, $\psi_{m,n}$ is the amplitude of the $m$th (right) eigenstate at site $n$ and normalized such that $\sum_{n} \abs{\psi_{m,n}}^2=1$.
A second defining spectral property is that the spectrum is real, and remains essentially insensitive to boundary conditions: switching between open (OBC) and periodic (PBC) boundary conditions does not induce the qualitative spectral reshaping characteristic of the conventional NHSE.
For comparison, Fig.~\ref{fig:schematic}d sketches the Anderson-localized situation.
% Here the eigenstates are exponentially localized around random centers throughout the system, producing many localized profiles and, consequently, a comparatively featureless mode-summed profile $\Psi_n$.
% In the Anderson model, the spectrum is also real by Hermiticity.
% Thus, at the level of coarse spectral attributes---real eigenenergies and spectrally localized eigenstates---ENHSL and Anderson localization may appear similar.

A major finding is that the dynamical response provides the sharpest contrast between ENHSL and Anderson localization.
Figure~\ref{fig:schematic}e illustrates the ENHSL dynamics following a bulk-centered excitation.
In a single disorder realization (3D traces), the wave packet does not simply spread and then freeze, nor does it become pinned to a static defect-like site.
Instead, the dynamics is dominated by an attraction toward a realization-dependent bulk region associated with the dominant excursion of the random-walk landscape $X_n$ within the finite-time light cone of the underlying disorder-free chain.
Accordingly, the wave-packet develops a pronounced peak that is strong, macroscopic, and highly sample specific.
Most importantly, ENHSL exhibits an ensemble-level transport behavior that is counterintuitive from the viewpoint of conventional disorder physics.
The ensemble statistics reveal that the spreading distance can remain ballistic on average, which is schematically indicated in Fig.~\ref{fig:schematic}e.
% This is schematically indicated in Fig.~\ref{fig:schematic}e, where the dashed guidelines $n=\pm\langle d(t)\rangle$ track the symmetric ensemble-averaged spreading envelope, and the gray shading indicates realization-to-realization fluctuations around this average.
In other words, ENHSL fully decouples spectral eigenstate localization from transport suppression, allowing ballistic spreading on average.
Figure~\ref{fig:schematic}f summarizes the Anderson-localized dynamical benchmark, where the wave packet becomes dynamically localized. %: after an initial transient, the probability distribution remains confined within a finite region set by the localization length, and the spreading distance saturates at long times.
Ensemble averaging does not restore ballistic motion; instead, it further reinforces the picture of inhibited transport typical of Anderson localization.

\subsection{Unified spectral-and-dynamical measurement protocol}
\begin{figure}[p]
    \centering
    \phantomsection 
    \includegraphics[width=0.95\textwidth]{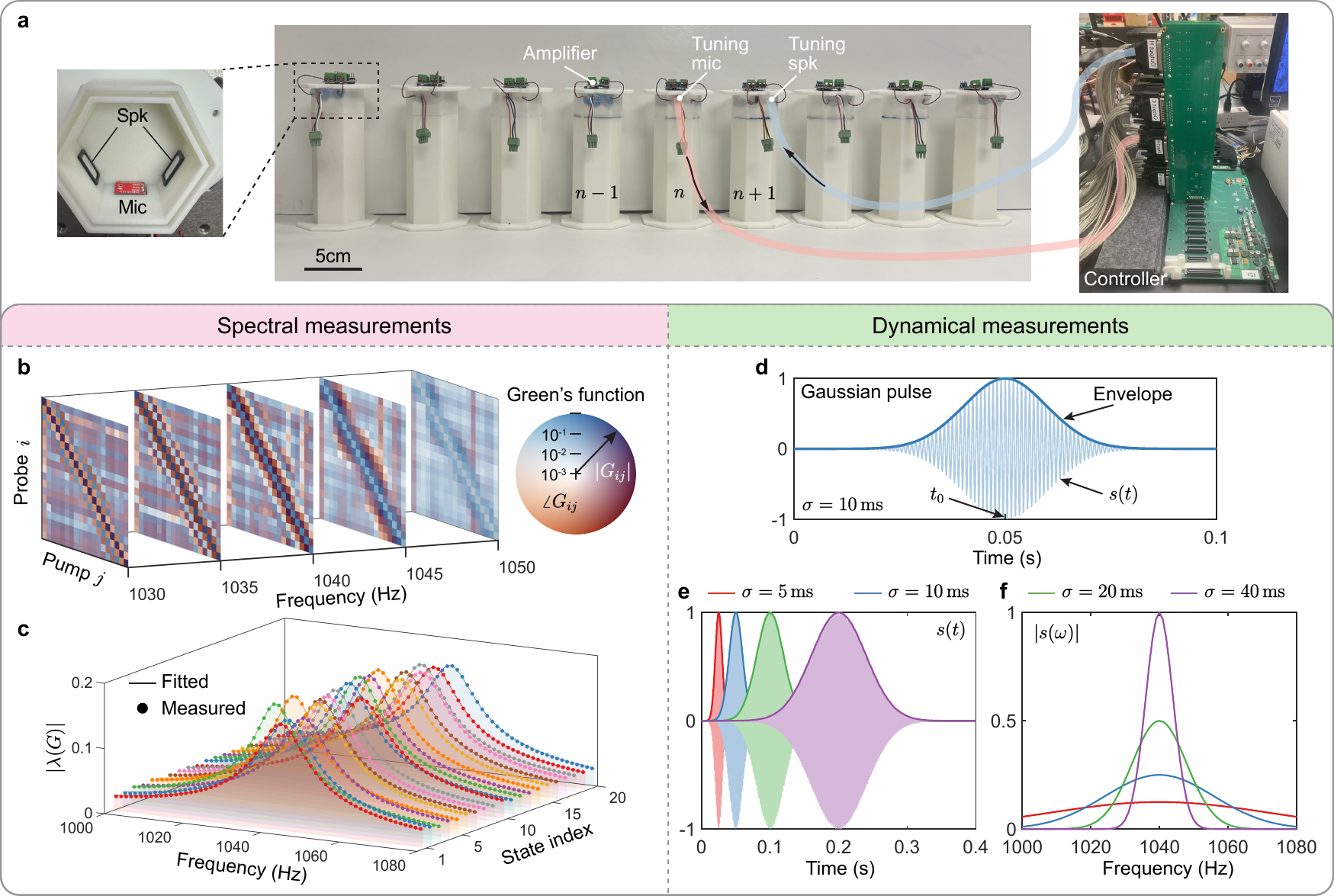} 
    \caption{\textbf{Unified spectral-and-dynamical measurement protocol on the same acoustic lattice.}
    \textbf{a}, Photograph of the experimental platform: a 1D chain of acoustic resonators (interior view shown at left, containing a microphone and two loudspeakers) coupled by electroacoustic feedback implemented with external amplifiers and phase shifters integrated in a multi-channel controller (right). 
    Here, a representative hopping from site $n$ to site $n+1$ is illustrated.
    \textbf{b,c}, Green's-function-based spectroscopy for full spectral reconstruction.
    \textbf{b}, Site-by-site excitation (pump $j$) together with full spatial readout (probe $i$) yields the frequency-resolved response (Green's-function) matrix $G_{ij}(\omega)$. 
    Shown is an example lattice of size $N=20$ under OBC ($\omega_0 = 1040\,\mathrm{Hz} - 6i\,\mathrm{Hz}$, $J=1.5\,\mathrm{Hz}, \Delta h = 2\,\mathrm{Hz}$), with $G_{ij}(\omega)$ visualized at several representative frequencies.
    The inset color wheel indicates the visualization scheme: hue encodes the phase $\angle G_{ij}$, while brightness encodes the magnitude $|G_{ij}|$ (log-scaled).
    \textbf{c}, Extraction of complex eigenenergies from the measured Green’s function. Dots show the measured magnitudes of the eigenvalues $\lambda_m(\omega)$ of $G(\omega)$ as functions of frequency, and solid curves show fits to a single-pole form $\lambda_m(\omega)\simeq A_m/(\omega-E_m)$, from which the complex eigenenergies $E_m$ are obtained. 
    \textbf{d--f}, Time-resolved wave-packet measurements using the same lattice.
    \textbf{d}, Gaussian-modulated tone burst $s(t)$ centered at $\Re (\omega_0)=1040\,\mathrm{Hz}$, illustrating the carrier and the Gaussian envelope (width $\sigma$) with temporal center $t_0 = 5\sigma$.
    \textbf{e,f}, Excitation waveforms $s(t)$ (e) and their spectra $|s(\omega)|$ (f) for four representative envelope widths $\sigma=5,\,10,\,20,$ and $40\,\mathrm{ms}$.
    }
    \label{fig:exp_setup}
    \addcontentsline{toc}{subsection}{Figure \thefigure}
\end{figure}

In this work, we experimentally realize the globally reciprocal Hatano-Nelson model in an engineered acoustic lattice platform \cite{Zhang2021AcousticNonHermitianSkin, Liu2022ExperimentalRealizationWeyl, Chen2025DirectMeasurementTopological, Tong2025ObservationFloquetBlochBraids, Zhong2025HigherorderSkinEffect, Zhong2025ExperimentallyProbingNonHermitian}. %, enabling unified spectral-and-dynamical measurements.
Spatial disorder is introduced in a controlled manner by implementing the link asymmetry according to a prescribed sequence $\{h_n\}$.
Different disorder realizations are generated by setting distinct sequences $\{h_n\}$, implemented through calibrated gain and phase settings in the amplifiers and phase shifters (see Methods).
We implement a unified spectral-and-dynamical protocol on the same acoustic lattice (Fig.~\ref{fig:exp_setup}a).
In the spectral measurements, we perform site-by-site excitation (pump at site $j$) while recording the steady-state responses at all sites (probe at site $i$), thereby obtaining the frequency-resolved Green's-function matrix $G_{ij}(\omega)$ (Fig.~\ref{fig:exp_setup}b).
% Diagonalizing $G(\omega)$ yields eigenvalue trajectories $\lambda_m(\omega)$ as functions of the drive frequency ($\omega$).
% We fit each trajectory to a single-pole form, $\lambda_m(\omega)\simeq {A_m}/({\omega-E_m})$, to extract the complex eigenenergies $E_m$ (Fig.~\ref{fig:exp_setup}c).
% The corresponding eigenstates are reconstructed from the fitted Green's-function response near $\omega\simeq \Re(E_m)$, equivalently from the eigenvectors of $G(\omega)$ near the resonance.
By diagonalizing the full measured Green's function matrix (Fig.~\ref{fig:exp_setup}c), we obtain the complete complex spectrum $\{E_m\}$ and corresponding eigenstates $\{\psi_{m,n}\}$, which further enable the evaluation of mode-resolved quantities such as the inverse participation ratio (IPR) and mode-summed intensity $\Psi_n$ (see \cite{Zhong2025ExperimentallyProbingNonHermitian} for details).
% Because the full matrix $G_{ij}(\omega)$ is measured, this procedure yields a comprehensive spectral characterization within the experimentally accessible band, including mode-resolved quantities such as the inverse participation ratio (IPR) and mode-summed intensity $\Psi_n$.
% See \cite{Zhong2025ExperimentallyProbingNonHermitian} for details of the spectral reconstruction procedure.

In the dynamical measurements, we excite the lattice with a Gaussian-modulated tone burst $s(t)$ centered at frequency $\Re (\omega_0)$ (Fig.~\ref{fig:exp_setup}d)
\begin{equation}
    s(t)=
    \exp\qty[{-\frac{(t-t_0)^2}{2\sigma^2} }-{i} \Re\qty(\omega_{0})t] 
    .
    \label{eq:gaussian_pulse}
\end{equation}
Here, $t_0 $ is the pulse center time, and $\omega_0$ is the complex resonant frequency of a single cavity with the imaginary part accounting for intrinsic losses. 
The pulse width $\sigma$ is chosen to satisfy two experimental constraints:
(i) $\sigma$ cannot be too small, otherwise the injected energy becomes insufficient and the propagating wave packet rapidly decays into the noise floor;
(ii) $\sigma$ cannot be too large, otherwise the excitation spectrum becomes too narrow to cover the relevant bandwidth of the lattice spectrum (Fig.~\ref{fig:exp_setup}e,f).
The time-resolved signals are simultaneously recorded at all sites using synchronized data acquisition modules. 
Importantly, the evenlope of these time-domain measurements are extracted to represent the wavefunctions $\psi_n(t)$ (see Methods).

% Crucially, the spectral and dynamical measurements are performed under the same hardware configuration and the same disorder realization, enabling direct cross-validation:
% the reconstructed eigenstates and eigenenergies provide the spectral signature of ENHSL, while the time-domain measurements directly reveal the corresponding propagation and ensemble transport statistics.
% Experimentally, we implement the chain using an active acoustic lattice in which electroacoustic feedback realizes tunable hoppings between neighboring sites as well as onsite potential adjustment \cite{Zhong2025ExperimentallyProbingNonHermitian} (see Methods).
% Spatial disorder is introduced in a controlled manner by implementing the link asymmetry according to a prescribed sequence $\{h_n\}$.
% Different disorder realizations are generated by setting distinct sequences $\{h_n\}$, implemented through calibrated gain and phase settings in the amplifiers and a customized controller (see Methods).

\subsection{Experimental observation of ENHSL spectral signatures}

\begin{figure}[p]
    \centering
    \phantomsection 
    \includegraphics[width=0.65\textwidth]{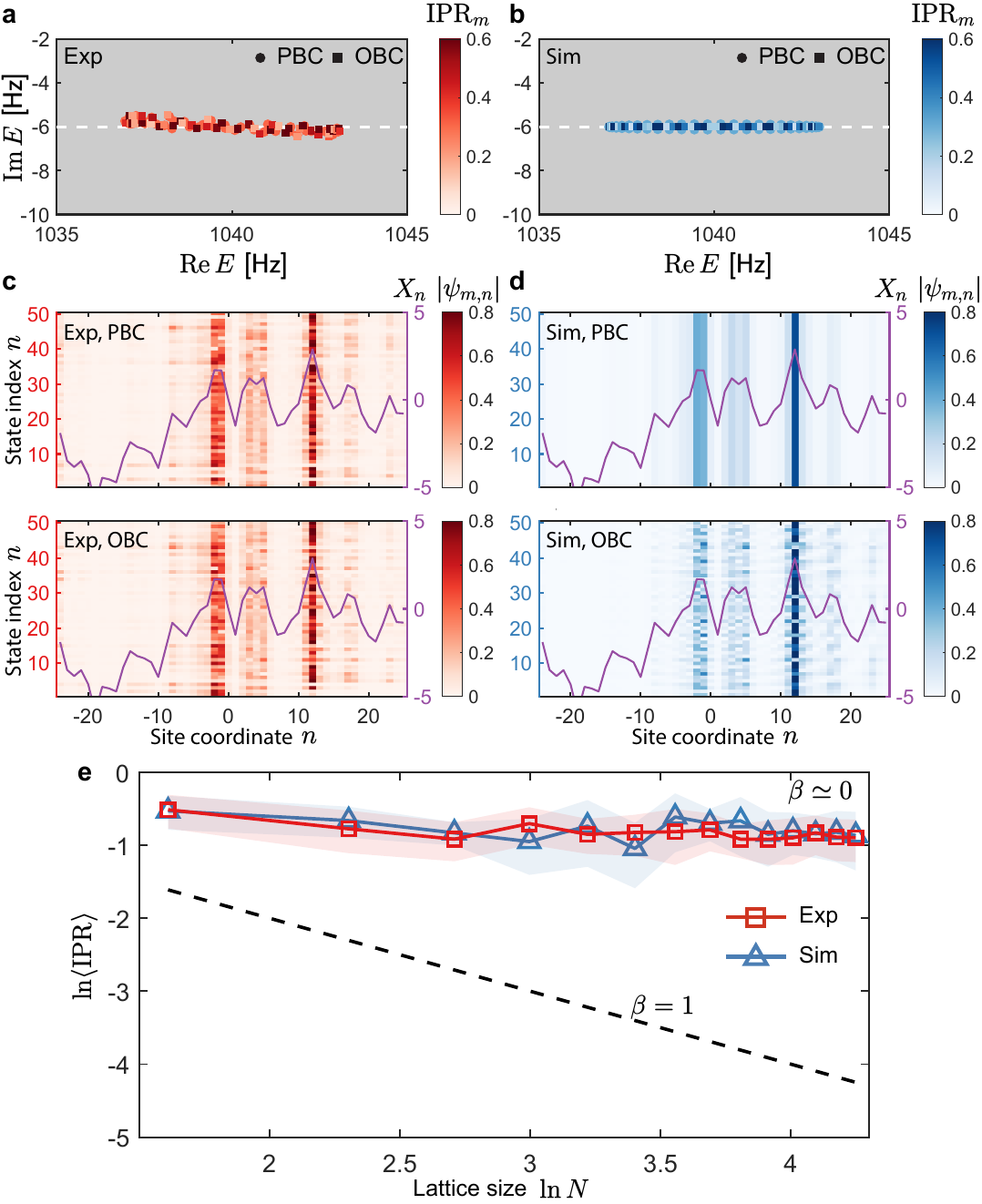}
        \caption{
            \textbf{Spectral signatures of ENHSL, comparing experiments and tight-binding simulations for a chain of size $N=50$ with $\omega_0 = 1040\,\mathrm{Hz} - 6i\,\mathrm{Hz}$, $J=1.5\,\mathrm{Hz}$, and $\Delta h = 2\,\mathrm{Hz}$.} 
            \textbf{a,b}, Complex spectra for a representative realization under both PBC and OBC. Circles (squares) denote PBC (OBC) energies, and colors encode $\mathrm{IPR}_m$. The dashed horizontal line marks the nominal uniform background-loss level ($-6\,\mathrm{Hz}$) used as a reference. 
            \textbf{c,d}, Spatial distributions of all eigenstates under PBC (top) and OBC (bottom) for experiment (c) and simulation (d). Heat maps show the eigenstate amplitudes $|\psi_{m,n}|$ as functions of eigenstate index $m$ and site coordinate $n$. The overlaid purple curve (right axis) shows the random-walk landscape $X_n$ associated with the imaginary-gauge-field disorder, highlighting that prominent localization peaks occur near large excursions of $X_n$. 
            \textbf{e}, Scaling of localization with system size under OBC. 
            Symbols show the ensemble-averaged $\langle{\mathrm{IPR}}\rangle$ for $N=5,10,15,...,70$ (each $N$ averaged over $R=8$ disorder realizations), comparing experiment and simulation; shaded bands indicate the realization-to-realization spread (standard deviation). 
            The dashed guideline ($\beta=1$) indicates the extended-state scaling $\mathrm{IPR}\propto N^{-1}$.
            A fit to $\langle{\mathrm{IPR}}\rangle \propto N^{-\beta}$ yields $\beta\simeq 0$, consistent with size-independent localization in ENHSL.
            }
    \label{fig:spectral}
    \addcontentsline{toc}{subsection}{Figure \thefigure}
\end{figure}

We first experimentally observe the spectral features of ENHSL under both PBC and OBC.
Figures~\ref{fig:spectral}a,b compare experimental and tight-binding simulated spectra for a representative realization, showing good agreement.
Despite locally non-reciprocal hoppings, the spectrum remains nearly real up to an overall uniform loss shift ($-6\,\mathrm{Hz}$) for both PBC and OBC.
% Colors encode $\mathrm{IPR}_m\equiv \sum_{n}\abs{\psi_{m,n}}^{4}$, which quantifies eigenstate localization.
Figures~\ref{fig:spectral}c,d show the spatial distributions of the eigenstates $|\psi_{m,n}|$, with the random-walk landscape $X_n$ overlaid (right axis).
ENHSL manifests as macroscopic localization peaks whose positions correlate with large excursions of $X_n$, and which persist and are only weakly modified when switching between PBC and OBC.
This weak boundary sensitivity contrasts sharply with the conventional NHSE in clean non-reciprocal lattices, where switching boundary conditions qualitatively reshapes both spectrum and localization \cite{Zhong2025ExperimentallyProbingNonHermitian}.
From Figs.~\ref{fig:spectral}b and \ref{fig:spectral}d, we see that all eigenstates have the same IPR under PBC, whereas under OBC a small subset of states shows noticeable deviations.
Nevertheless, all eigenstates remain strongly localized under both boundary conditions.

To quantify the system-size dependence, we performed experiments for varying lattice size $N=5$ to $N=70$ with a step $5$ under OBC.
For each lattice size, we measured $R=8$ disorder realizations (see Supplementary Materials for representative realizations).
For each realization, we compute the mode-averaged IPR, $\overline{\mathrm{IPR}}_r=N^{-1}\sum_{m=1}^{N}\mathrm{IPR}_m$, and then average over realizations to obtain $\langle{\mathrm{IPR}}\rangle=R^{-1}\sum_{r=1}^{R}\overline{\mathrm{IPR}}_r$.
Figure~\ref{fig:spectral}e presents the averaged IPR versus chain length $N$, compared between experiment and simulation showing good agreement.
% The fractal dimension $\beta$ is defined by $\beta = -\ln(\overline{\mathrm{IPR}})/\ln N$, characterizing the scaling of the IPR with system size.
We characterize the size scaling by fitting $\langle{\mathrm{IPR}}\rangle \propto N^{-\beta}$, where $\beta$ is an effective scaling exponent.
For extended and localized states, $\beta$ takes the values $1$ and $0$, respectively, while $0<\beta<1$ indicates fractal states.
% It is seen from Fig.~\ref{fig:spectral}e that $\ln \qty(\overline{\mathrm{IPR}})$ remains approximately constant as $\ln N$ increases, yielding $\beta \approx 0$.
As shown in Fig.~\ref{fig:spectral}e, $\ln\langle{\mathrm{IPR}}\rangle$ is approximately independent of $\ln N$, yielding $\beta\simeq 0$.
This experimental observation confirms the strong spectral localization of eigenstates in ENHSL, although the localization is sub-exponential due to the vanishing of the Lyapunov exponent \cite{Longhi2025ErraticNonHermitianSkin}.

\subsection{Theoretical description of ENHSL dynamics}
\label{sec:dynamics_theory_main}

The spectral formulation of ENHSL describes eigenstate localization via the random-walk landscape $X_n$ induced by the stochastic imaginary gauge field.
Here, we present a complementary dynamical perspective. %, which yields two experimentally testable predictions:  
% (i) in a single disorder realization the evolving wave packet is attracted toward a realization-dependent dominant site set by the extremal statistics of $X_n$ within a finite-time light cone; and
% (ii) despite spectral localization, ensemble-averaged transport remains ballistic, with the dominant-site distribution obeying a universal L\'evy-arcsine law.
% 
To reveal the single-particle dynamics, we apply the non-unitary gauge transformation $\psi_n(t) = \phi_n(t)\, \exp(X_n)$, which maps the system onto a uniform Hermitian nearest-neighbor chain with hopping $J$. For a single-site initial excitation at $n=0$, the Hermitian solution is given by $\phi_n(t) = (-i)^n J_n(2J t)$, where $J_n(\cdot)$ is the Bessel function of first kind.
Transforming back to the original variables, the normalized occupation probability profile at site $n$ and time $t$ is
\begin{equation}
    P_n(t)=\frac{|\psi_n(t)|^2}{\sum_{n'}|\psi_{n'}(t)|^2}
    =\frac{J_n^2(2J t)\exp\qty(2X_n)}{\sum_{n'} J_{n'}^2(2J t)\exp\qty(2X_{n'})}.
\end{equation}
We quantify transport in a single realization using the second moment $d(t)=\sqrt{\sum_n n^2 P_n(t)}$.
In the disorder-free limit $h_n=0$ (hence $X_n=0$), one recovers ballistic spreading $d(t)=\sqrt{2}\,J t$.
Therefore, the disorder-free ballistic spreading is retained in the underlying propagator, while the stochastic gauge factors reshape where intensity accumulates.
As a result, the finite-time intensity profile can be viewed as a Hermitian ballistic core modulated by an envelope $\exp\qty(2X_n)$, which locally amplifies ($X_n>0$) or suppresses ($X_n<0$) intensity.

At finite time $t$, the Bessel kernel confines the dynamics with a super-exponential localization to a light-cone-like accessible interval
$|n|\lesssim n_{\mathrm{LB}}(t)$, with a linear-in-time Lieb-Robinson (LB) bound $n_{\mathrm{LB}}(t)\simeq 2Jt$.
Within this interval, ENHSL dynamics is controlled by an extremal statistic:
the wave packet is predominantly localized around the dominant site
$ n_0(t)=\arg\max_{n} P_n(t),$
which follows the location where $X_n$ attains its largest excursion within the light cone.
This dominant-site picture is crucial: the dynamics is governed by the maximum excursion of a symmetric random walk
over an expanding interval, rather than by typical disorder fluctuations.
As $X_n$ is a symmetric random walk, the position of its maximum over a finite interval exhibits a universal limiting
distribution according to the Sparre-Anderson theorem \cite{Andersen1953FluctuationsSumsRandom, Andersen1954FluctuationsSumsRandom, Spitzer1964PrinciplesRandomWalks, Feller1968IntroductionProbabilityTheory}.
Consequently, for sufficiently large $t$ the empirical distribution of $n_0(t)$ approaches the L\'evy--arcsine law mapped
to the interval $[-n_{\mathrm{LB}}(t),\,n_{\mathrm{LB}}(t)]$ (derivation in Methods),
\begin{equation}
P_{\mathrm{arc}}(n_0,t)=\frac{1}{\pi\sqrt{\bigl[n_{\mathrm{LB}}(t)-n_0\bigr]\bigl[n_0+n_{\mathrm{LB}}(t)\bigr]}},
\qquad |n_0|<n_{\mathrm{LB}}(t).
\label{eq:arcsine_main}
\end{equation}
Equation~\eqref{eq:arcsine_main} predicts an enhanced probability of finding the dominant site near the light-cone edges,
$|n_0|\sim n_{\mathrm{LB}}(t)$.
Furthermore, by using Eq.~\eqref{eq:arcsine_main}, the second moment averaged over all locations can be approximated as $\langle d(t)\rangle \simeq \sqrt{\sum_n n^2 P_\mathrm{arc}(n_0=n, t) } = \sqrt{2}Jt$, which clearly shows the ballistic scaling.
% Since the instantaneous profile $P_n(t)$ is sharply peaked around $n_0(t)$ in the ENHSL regime, this yields ballistic scaling for propagation-range observables, consistent with the measured ensemble behaviour discussed below.

\subsection{Experimental observation of single-realization dynamics}
\label{sec:single_realization}

Guided by the dynamical theory, we first experimentally observe the extremal-selection mechanism at the level of individual disorder realizations.
Figures~\ref{fig:single_realization_dynamics}a--c presents three representative realizations for ENHSL dynamics (see Supplementary Materials for more realizations), compared with two reference cases: (Fig.~\ref{fig:single_realization_dynamics}d) clean Hermitian chain (ballistic spreading benchmark) and (Fig.~\ref{fig:single_realization_dynamics}e) a disordered Hermitian chain (Anderson localization benchmark).
For each ENHSL realization (Fig.~\ref{fig:single_realization_dynamics}a--c), the mode-summed eigenstate intensity $\Psi_n = \sum_m \abs{\psi_{m,n}}^2$ exhibits a prominent macroscopic peak (often accompanied by a few weaker satellite peaks).
This peak structure is fundamentally different from the Anderson reference (Fig.~\ref{fig:single_realization_dynamics}e), where the sum over many exponentially localized eigenstates with random centers yields a comparatively featureless $\Psi_n$.
In ENHSL, the random-walk profile $X_n$ shown in the second row of each panel acts as the organizing structure: prominent spectral peaks emerge near its large excursions, consistent with the spectral picture.
% The key dynamical question is then how the time-resolved wave packet $\psi_n(t)$ responds to this underlying structure.

The third row of Fig.~\ref{fig:single_realization_dynamics} shows the measured spatiotemporal evolution $\abs{\psi_n(t)}$ following excitation at the central site.
The corresponding simulation results are presented in the fourth row, showing good agreement with experiment for all configurations.
In each ENHSL realization, $\abs{\psi_n(t)}$ develops a pronounced localization peak at an irregular bulk position, in agreement with the theoretical picture that dynamics is drawn toward a dominant excursion of $X_n$.
Importantly, this localization peak does not simply coincide with a fixed Anderson-like center; rather, the region of maximal intensity is selected by the disorder realization and can evolve in time within the expanding light cone interval.
In several realizations, the dominant intensity ridge in the $(n,t)$ plane reveals a slow drift of the peak position as time increases, reflecting the fact that $n_0(t)$ is defined by an extremal statistic on an expanding interval $\qty(-n_\mathrm{LB}(t), n_\mathrm{LB}(t))$.
In constrast, the Hermitian disorder-free reference (Fig.~\ref{fig:single_realization_dynamics}d) exhibits the ballistic spreading pattern, while the Anderson reference (Fig.~\ref{fig:single_realization_dynamics}e) shows rapid confinement of $\abs{\psi_n(t)}$ near the initial excitation site, consistent with dynamical localization.

To quantify spreading, we extract a spreading distance $d(t)$ from $\abs{\psi_n(t)}$ and compare experiment and tight-binding simulations (bottom row of Fig.~\ref{fig:single_realization_dynamics}).
% Again, good agreement is seen across all configurations.
For ENHSL realizations, $d(t)$ can grow fast and saturate slowly (Fig.~\ref{fig:single_realization_dynamics}a) or grow slower than the ballistic reference (Fig.~\ref{fig:single_realization_dynamics}c), or exhibit intermediate behaviours (Fig.~\ref{fig:single_realization_dynamics}b).
These representative realizations show distinct evolution patterns for ENHSL depending on the disorder realization, reflecting strong sample-to-sample fluctuations.
For reference, the clean Hermitian chain (Fig.~\ref{fig:single_realization_dynamics}d) shows $d(t)$ growing almost linearly with time, while the Anderson reference (Fig.~\ref{fig:single_realization_dynamics}e) shows $d(t)$ saturating quickly.
% Despite the strong realization-to-realization variability, a common feature of all ENHSL realizations is that $d(t)$ grows significantly over time, in stark contrast to the Anderson case.

% This coexistence is absent in the Anderson reference (Fig.~\ref{fig:single_realization_dynamics}e), where $P_n(t)$ remains confined near the excitation region and $d(t)$ saturates, indicating dynamical localization.

\subsection{Experimental observation of ensemble dynamics and L\'evy-arcsine law}

\begin{figure}[]
    \centering
    \phantomsection 
    \includegraphics[width=0.99\textwidth]{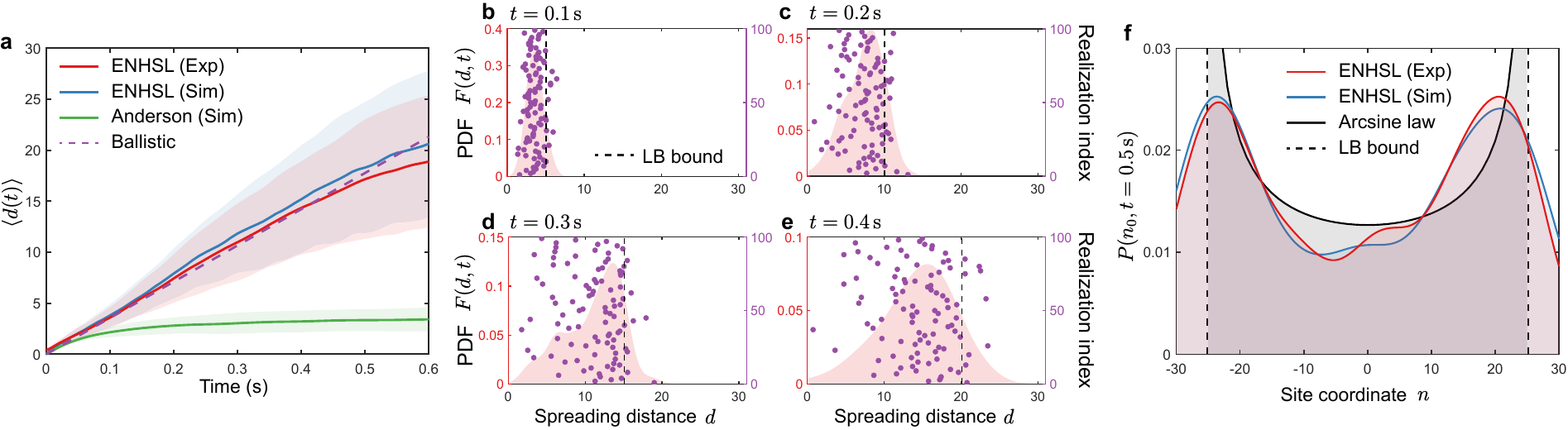}
    \caption{\textbf{Ensemble statistics of wave-packet spreading for the ENHSL (OBC, $N=61$), obtained from $R=100$ disorder realizations. }
    The parameters set in experiments are \(\omega_0 = 1040\,\mathrm{Hz} - 1.75i\,\mathrm{Hz}\), \(J=4\,\mathrm{Hz}\), and \(\Delta h = 1.6\,\mathrm{Hz}\).
    \textbf{a}, Ensemble-averaged spreading distance $\langle d(t)\rangle$ from experiments and tight-binding simulations, compared with the ballistic scaling of a clean Hermitian chain and with a Hermitian Anderson-localized reference (shown by simulations, exhibiting saturation). 
    \textbf{b--e}, Distributions of the spreading distance at fixed times $t=0.1,\,0.2,\,0.3,$ and $0.4~\mathrm{s}$, respectively. In each panel, purple markers indicate the values $d^{(r)} (t)$ from individual realizations $r$ (realization index shown on the right axis). The shaded curves show the probability density function (PDF) $F(d,t)$ (left axis) estimated from the finite ensemble using a bounded KDE method on the interval $(0,30)$. The dashed lines denote the Lieb-Robinson (LB) bound, $n_\mathrm{LB}(t) = 2Jt$. 
    \textbf{f}, PDF of the dominant propagation site, $n_0(t)=\arg\max_n \abs{\psi_n(t)}$, at $t=0.5~\mathrm{s}$, obtained from the same $R=100$ realizations, and estimated using a bounded KDE method on the interval $(-30,30)$. Experimental and simulated distributions are compared. The dashed vertical lines indicate the corresponding LB bounds, $\pm n_{\mathrm{LB}}$, and the black curve shows the L\'evy arcsine-law prediction restricted to the interval $(-n_{\mathrm{LB}},\,n_{\mathrm{LB}})$.}
    \label{fig:statistics}
    \addcontentsline{toc}{subsection}{Figure \thefigure}
\end{figure}

We now turn from representative realizations to ensemble-level transport and statistics. %, where the dynamical theory makes two quantitative predictions.
Figure~\ref{fig:statistics}a shows the ensemble-averaged ($R = 100$ realizations) spreading distance $\langle d(t)\rangle $ extracted from the measured wave packets and compared with tight-binding simulations.
The ENHSL data exhibit a robust linear growth over the experimentally accessible time window, consistent with ballistic transport.
For comparison, the clean Hermitian chain shows the ballistic scaling, whereas the Anderson-localized reference saturates, reflecting conventional dynamical localization.
% That ENHSL follows the ballistic trend is remarkable because, as seen in Fig.~\ref{fig:single_realization_dynamics}, individual realizations are strongly concentrated around a dominant peak.
% The ensemble result therefore directly supports the central claim of the dynamics theory: ENHSL decouples strong spectral localization from the ballistic transport.

Beyond the mean, Figs.~\ref{fig:statistics}b--e present the distributions of $d(t)$ at several fixed times.
To faithfully estimate the probability density from a finite ensemble without boundary artefacts, we employ a bounded kernel density estimation (KDE) on the finite support of $d$ (caption of Fig.~\ref{fig:statistics}).
% The same estimation procedure is applied to experimental and simulated datasets to enable a controlled comparison.
These panels visualize two important aspects of ENHSL dynamics.
First, the distributions are broad and strongly time dependent, reflecting significant realization-to-realization fluctuations.
This broadness is expected because the dynamics is governed by extremal statistics of the random walk $X_n$, which naturally produces large sample-to-sample variability.
Second, the weight of the distribution shifts to larger $d$ as time increases, consistent with the expansion of the accessible light-cone interval.
In contrast, the Anderson reference would exhibit distributions that rapidly narrow and remain confined, consistent with saturation of $d(t)$.

While $d(t)$ provides a natural bulk measure of spreading, the most discriminating test of the dynamics theory concerns the dominant-site statistics.
For each realization, we extract $n_0(t)=\arg\max_{n} \abs{\psi_n(t)}$, and obtain its distribution $P(n_0,t)$.
The theory predicts that $n_0(t)$ follows the Lévy--arcsine law on the finite interval $(-n_{\mathrm{LB}}(t),\,n_{\mathrm{LB}}(t))$, Eq.~\eqref{eq:arcsine_main}, where the bound $n_{\mathrm{LB}}(t) = 2Jt$ is fixed by the independently known coupling $J$.
Figure~\ref{fig:statistics}f shows the measured $P(n_0,t)$ at $t=0.5~\mathrm{s}$ (correspondingly, $n_\mathrm{LB} = 25$) and compares it with both simulation and the arcsine prediction.
The measured distribution is strongly nonuniform and enhanced toward the edges of the accessible interval, as expected from the arcsine law.
This agreement constitutes a stringent verification because it tests the full functional form rather than just a scaling trend, and hinges on an extremal statistic that is explicitly connected to the underlying random walk.

The arcsine-law verification also provides an explanation for ballistic ensemble transport.
A symmetric random walk attains its maximum near the boundaries of a finite interval with high probability, and therefore the typical magnitude of the maximizer scales as $|n_0|\sim n_{\mathrm{LB}}(t)$.
Since $n_{\mathrm{LB}}(t)\propto t$, the typical displacement associated with the dominant peak grows linearly, yielding ballistic scaling for $\langle d(t)\rangle$ even though each realization appears localized at any fixed time.

\section{Discussions}
Our results provide the first experimental demonstration of a fundamentally new localization regime in non-Hermitian systems, in which strong, macroscopic localization robustly coexists with ballistic transport. 
Distinct from both Hermitian Anderson localization---where spectral localization is inseparable from dynamical arrest---and from the conventional NHSE---where bulk modes accumulate exponentially at boundaries---ENHSL establishes a fundamentally different paradigm: globally reciprocal non-Hermiticity and disorder jointly generate realization-dependent bulk localization peaks with subexponential profiles, while the associated wave-packet dynamics remains transport preserving on average.
A central outcome of our work is the identification and verification of an extremal-statistics mechanism for transport. By combining full spectral reconstruction with time-resolved measurements on the same platform, we directly connect the dominant propagation site to the largest excursion of a symmetric random-walk landscape within a finite-time light cone, and we confirm the resulting universal L\'evy-arcsine statistics. 
Beyond identifying a new form of wave localization, our findings open new directions in the study of complex wave dynamics and reveal a broadly applicable mechanism relevant to quantum and classical waves alike---from condensed matter and photonics to acoustics, metamaterials and other complex media.
More generally, ENHSL highlights a route toward engineering stochastic yet controllable wave confinement with predictable ensemble transport, and it motivates future studies of how extremal statistics, topology, interactions, or higher dimensions reshape the relation between localization and propagation in complex non-Hermitian systems.

\section{Acknowledgment}
Y. J. thanks the support of startup funds from Penn State University and NSF CMMI awards 2039463 and 195122.

\section{Author contributions}

J.-X. Z., S. L., and Y. J. conceived the project.
S. L. and J.-X. Z. performed theoretical analysis and numerical simulations.
J.-X. Z. designed and performed the experiments with assistance from J. W. K..
J.-X. Z., S. L., and Y. J. wrote the paper.
S. L. and Y. J. supervised the project.

\section{Competing interests}
The authors declare no competing interests.

\clearpage
\bibliographystyle{unsrt}
\bibliography{bibtex}
% \printbibliography
\addcontentsline{toc}{section}{References}

\clearpage 
% ============================================================
% Methods 
% ============================================================
\section{Methods}

\subsection{Details of dynamics theory of ENHSL}
\label{sec:methods_dynamics_theory}

\paragraph{Model and evolution equation.}
We consider the 1D Hatano-Nelson chain with stochastic imaginary gauge field $h_n$ on each link, with hoppings shown in Eq.~\eqref{eq:HN_hoppings}.
In the single-particle sector, the dynamical equations for the (non-normalized) probability amplitudes $\psi_n(t)$ satisfy
\begin{equation}
i\frac{d\psi_n}{dt}
=
J^{\mathrm{L}}_{n}\psi_{n+1}+J^{\mathrm{R}}_{n-1}\psi_{n-1}
=
J\left(e^{-h_n}\psi_{n+1}+{e}^{h_{n-1}}\psi_{n-1}\right),
\label{eq:eom_methods}
\end{equation}
with either OBC $(\psi_0 = \psi_{N+1} = 0)$ or PBC $(\psi_{n+N}=\psi_n)$.
Note that under OBC, boundary effects play no role in wave packet expansion provided the excitation starts in the bulk and the wave packet does not reach the edges within the observation time.

\paragraph{Random-walk landscape and non-unitary gauge transformation.}
The cumulative gauge field (random-walk landscape) is defined as Eq.~\eqref{eq:random_walk_landscape}, which maps the sequence $\{h_n\}$ to a discrete-time symmetric random walk in $n$.
Introduce the non-unitary gauge transformation
\begin{equation}
\psi_n(t)=\phi_n(t)\,\exp(X_n)\equiv \phi_n(t)\,u_n,
\qquad u_n=\exp(X_n).
\label{eq:gauge_methods}
\end{equation}
Substituting Eq.~\eqref{eq:gauge_methods} into Eq.~\eqref{eq:eom_methods} yields the uniform Hermitian transport equation
\begin{equation}
i\frac{d\phi_n}{dt}=J(\phi_{n+1}+\phi_{n-1}).
\label{eq:hermitian_methods}
\end{equation}
For a single-site initial excitation at $n=0$, the Hermitian solution is
\begin{equation}
\phi_n(t)=(-i)^n J_n(2J t),
\label{eq:bessel_methods}
\end{equation}
where $J_n(\cdot)$ is the Bessel function of the first kind.
Therefore,
\begin{equation}
|\psi_n(t)|^2
=
J_n^2(2J t)\,\exp(2X_n)
=
J_n^2(2J t)\,u_n^2.
\label{eq:psi2_methods}
\end{equation}
In the experiment we analyze the normalized occupation probability profile
\begin{equation}
P_n(t)=\frac{|\psi_n(t)|^2}{\sum_{n'}|\psi_{n'}(t)|^2}
=\frac{J_n^2(2J t)\exp(2X_n)}{\sum_{n'} J_{n'}^2(2J t)\exp(2X_{n'})}.
\label{eq:P_methods}
\end{equation}
Equation~\eqref{eq:P_methods} makes explicit the central structure.
The temporal spreading of the wave packet resembles the standard Hermitian Bessel profile, modulated by the random envelope $u_n^2 = \exp(2X_n)$, which locally amplifies ($X_n>0$) or suppresses ($X_n<0$) intensity.
The excitation is statistically drawn toward the dominant excursion (local maxima) of $X_n$ within the dynamically accessible interval, producing strong localization peaks in each realization.

\paragraph{Light-cone bound and unit convention.}
The Bessel kernel $J_n^2(2J t)$ decays rapidly, with tails falling faster than exponentially outside the interval $|n|\gtrsim 2J t$, effectively defining a Lieb-Robinson (LR) bound—or light cone—for nearest-neighbor hopping:
\begin{equation}
|n|\lesssim n_{\mathrm{LB}}(t),\qquad n_{\mathrm{LB}}(t)\simeq 2J t.
\label{eq:LB_methods}
\end{equation}
If the hopping is reported in angular-frequency units (rad/s), Eq.~\eqref{eq:LB_methods} is used directly.
If the hopping is reported in Hz (cycles/s), one should convert $J_{\omega}=2\pi J_{\mathrm{Hz}}$ and write
\begin{equation}
n_{\mathrm{LB}}(t)\simeq 2J_{\omega}t=2(2\pi J_{\mathrm{Hz}})\,t.
\label{eq:LB_methods_Hz}
\end{equation}
In this work, all hoppings and onsite potentials are reported in Hz, so Eq.~\eqref{eq:LB_methods_Hz} is used.

\paragraph{Second-moment spreading observable.}
The evolution of the wave packet's spatial spread in a single-realization of disorder is quantified using the second moment:
\begin{equation}
d(t)=\sqrt{\sum_n n^2 P_n(t)}
=
\sqrt{
\frac{\sum_n n^2 J_n^2(2J t)\exp(2X_n)}{\sum_n J_n^2(2J t)\exp(2X_n)}
}.
\label{eq:d_methods}
\end{equation}
In the disorder-free limit $h_n=0$ (hence $X_n=0$), one recovers ballistic spreading $d(t)=\sqrt{2}\,J t$.

\paragraph{L\'evy-arcsine law for dominant-site distribution.}
Let $u_n=\exp(X_n)$ and consider the thermodynamic limit $N \rightarrow \infty$. From the central limit theorem, for large $n$ one has
\begin{equation}
\frac{1}{\sqrt{2n}}\log\left|\frac{u_n}{u_0}\right|\sim \mathcal{N}(0,\Delta h^2),
\end{equation}
where $\mathcal{N}(0,\Delta h^2)$ is the normal distribution with zero mean and variance $\Delta h^2$, so that typical envelope amplitudes behave as
\begin{equation}
u_n\sim \exp\!\left(-\Delta h\sqrt{n}\right),
\label{eq:envelope_decay}
\end{equation}
showing a sub-exponential but stronger-than-algebraic decay \cite{Longhi2025ErraticNonHermitianSkin}.

For finite time $t$, the Bessel profile $J_n^2(2Jt)$ is confined within the light cone $|n|<n_\mathrm{LB}(t) = 2Jt$, with rapid decay outside.
Thus $|\psi_n(t)|^2$ is predominantly localized around the site $n=n_0(t)$ where $X_n$ attains its maximum within $(-n_\mathrm{LB}(t),n_\mathrm{LB}(t))$:
\begin{equation}
    n_0(t) = \arg \max_{-n_\mathrm{LB} <n<n_\mathrm{LB}} X_n.
    \label{eq:n0_definition}
\end{equation}
Here $n_0$ is a random variable with a universal limiting distribution for large $t$, which follows from the Sparre-Andersen
theorem~\cite{Andersen1953FluctuationsSumsRandom, Andersen1954FluctuationsSumsRandom, Spitzer1964PrinciplesRandomWalks, Feller1968IntroductionProbabilityTheory}.
This theorem states that many fluctuation properties of symmetric random walks do not depend on the actual distribution $f(h)$,
provided that increments are continuous and the distribution is symmetric about $h=0$.
Assuming that the random walk has independent and identically distributed symmetric $\pm h$ increments, the probability distribution can be calculated using
standard ballot/reflection arguments~\cite{Spitzer1964PrinciplesRandomWalks, Feller1968IntroductionProbabilityTheory} and reads
\begin{equation}
P(n_0=n,t)=
    \frac{\binom{2(n+n_\mathrm{LB})}{n+n_\mathrm{LB}}\binom{2(n_\mathrm{LB}-n)}{n_\mathrm{LB}-n}}{\binom{4n_\mathrm{LB}}{2n_\mathrm{LB}}},
\qquad
n=-n_\mathrm{LB}+1,\ldots,n_\mathrm{LB}-1,
\end{equation}
with $n_\mathrm{LB}(t)=2Jt$.
Using Stirling's approximation,
\begin{equation}
\binom{2n}{n}\sim \frac{4^n}{\sqrt{\pi n}},
\label{eq:stirling_approx}
\end{equation}
for large $n$, this reduces to
\begin{equation}
P_\mathrm{arc}(n_0=n,t)\sim \sqrt{\frac{n_\mathrm{LB}}{\pi (n-n_\mathrm{LB})(n+n_\mathrm{LB})}}.
\label{eq:arcsine_derivation}
\end{equation}
The resulting distribution of $n_0$ is the universal arcsin L\'evy law, which indicates an enhanced probability of finding
$n_0$, and thus localization of excitation, near the light-cone edges $n\to \pm n_\mathrm{LB}=\pm 2Jt$.

\paragraph{Second moment averaged over disorder realizations.}
An approximate expression of the second moment averaged over all locations can be computed as 
\begin{equation}
\langle d(t)\rangle = \sqrt{\sum_{n=-n_\mathrm{LB}}^{n_\mathrm{LB}} n^2 P_\mathrm{arc}(n_0=n, t) } 
% =
%     \sqrt{\sum_{n=-n_\mathrm{LB}}^{n_\mathrm{LB}} \frac{n^2}{\pi {n_\mathrm{LB}} \sqrt{(1-x)(1+x)}}}
.
\end{equation}
By setting $n = x n_\mathrm{LB}$ with $x\in[-1,1]$ and using Eq.~\eqref{eq:arcsine_derivation}, in the limit $n_\mathrm{LB} \to \infty$ one has 
\begin{equation}
\langle d(t)\rangle = n_\mathrm{LB} \sqrt{\int_{-1}^{1} x^2 P_0 (x) \dd x} = \sqrt{2}Jt,
\label{eq:average_d}
\end{equation}
where $P_0(x) = \frac{1}{\pi \sqrt{1-x^2}}$ is the standard arcsine distribution on $[-1,1]$, and the second moment of $P_0(x)$ is known to be $1/2$.
Equation~\eqref{eq:average_d} clearly shows the ballistic scaling of the disorder-averaged second moment.
In our experiments, the accessible sites extend to $|n|\sim 30$, for which the Stirling-based asymptotic approximation Eq.~\eqref{eq:stirling_approx} is already accurate at the percent level.

\subsection{Implementation of non-Hermitian acoustic lattices}

% Supplementary Fig. S1 shows a photograph of our experimental setup. 
The non-Hermitian acoustic lattices consists of multiple 3D-printed acoustic cavities (see Supplementary Materials for dimensions), each representing a site in the non-Hermitian lattice and operating near its first dipole resonance around 1040\,Hz.
The tuning of onsite potentials and hoppings is implemented using active components, specifically loudspeaker-microphone (pump-probe) pairs, an audio amplifier (LM386), and an phase shifter integrated in a customized controller (see Supplementary Materials for details).
The loudspeaker and microphone are positioned at the bottom of each cavity. 
The microphone captures the acoustic pressure signal, which is processed by the amplifier and phase shifter to adjust its amplitude and phase before being emitted by the loudspeaker in the connected cavity (see \cite{Zhong2025ExperimentallyProbingNonHermitian} for details).

% Nearest-neighbor hopping is realized by  the microphone signal from site $n$ to the loudspeaker at site $n\!+\!1$ (right hopping) and to the loudspeaker at site $n\!-\!1$ (left hopping) through digitally controlled gain and phase elements.

The imaginary gauge disorder is implemented by implementing an asymmetry between the right and left hopping magnitudes according to Eq.~\eqref{eq:HN_hoppings}, i.e., $J^{\mathrm{R}}_{n}/J^{\mathrm{L}}_{n}=\exp(2h_n)$, while keeping the geometric mean fixed at $J$.
Global reciprocity is enforced by choosing sequences $\{h_n\}$ satisfying $\sum_n h_n=0$.
In practice, the values are calibrated by independent measurements of the effective hoppings (e.g., via two-site characterization, see \cite{Zhong2025ExperimentallyProbingNonHermitian}).

\subsection{Spectral reconstruction via Green's function measurements}

For a chain of $N$ sites, we measure the steady-state complex response at all probe sites $i$ for each pump site $j$ under monochromatic excitation at frequency $\omega$.
This yields the full response matrix $G_{ij}(\omega)$ over a discrete set of drive frequencies spanning the band of interest.
The magnitude and phase are obtained by referencing the microphone signals to the excitation signal, allowing coherent extraction of $G_{ij}(\omega)$.
The full matrix acquisition over all pump-probe pairs yields $N^2$ complex transfer functions per frequency point \cite{Zhong2025ExperimentallyProbingNonHermitian}.

In the effective tight-binding description, the Green's function satisfies
\begin{equation}
G(\omega) = (\omega - H)^{-1},
\end{equation}
so the eigenvectors of $G(\omega)$ coincide with those of $H$, and its eigenvalues follow the single-pole form 
\begin{equation}
\lambda_m(\omega)=\frac{1}{\omega-E_m}.
\label{eq:pole_fit}
\end{equation}
up to a complex prefactor set by the measurement normalization.
We therefore diagonalize $G(\omega)$ for each sampled $\omega$, track each eigenvalue branch $\lambda_m(\omega)$ across frequency, and fit to Eq.~\eqref{eq:pole_fit} to extract $E_m$.
This procedure yields both $\Re(E_m)$ and $\Im(E_m)$, allowing direct access to the complex spectrum.
Eigenmodes are obtained from the corresponding eigenvectors of $G(\omega)$ evaluated near $\omega\simeq \Re(E_m)$, where the response is dominated by the target pole.
If needed, left eigenvectors can be obtained analogously from the left eigen-decomposition of $G(\omega)$; in the present work, we focus on the mode amplitudes and IPR used in Fig.~\ref{fig:spectral}.

\subsection{Time-resolved propagation measurements}

Time-domain dynamics are initiated by driving the central site ($n=0$) with a Gaussian-modulated tone burst,
Eq.~\eqref{eq:gaussian_pulse}. The carrier frequency is set to $\Re(\omega_0)=1040~\mathrm{Hz}$ (near the band center),
and the envelope width is chosen as $\sigma =10\,\mathrm{ms}$ following the criteria discussed in the main text
(Fig.~\ref{fig:exp_setup}d--f).
The driving signal is generated by a multi-channel data-acquisition card (PCIe-6353, National Instruments) and applied to
the on-site loudspeaker (CMS-15118D-L100, CUI Devices) through an audio amplifier (LM386, Texas Instruments).
The time-domain waveform at each site is recorded simultaneously using MEMS microphones (BOB-19389, SparkFun Electronics) at a
sampling rate of $16~\mathrm{kHz}$, with all channels acquired synchronously.

The recorded pressure signals are digitally bandpass-filtered around the carrier frequency (1010--1070~Hz).
The envelope of the signal is taken as $\abs{\psi_n(t)}$.
At each time $t$, we normalize $\psi_n(t)$ by its instaneous norm, so that $\sum_n \abs{\psi_n(t)}^2=1$ for all $t$.

To obtain ensemble statistics, we repeat the measurement for $R$ disorder realizations and evaluate $n_0^{(r)}(t)$ at a
fixed evolution time $t$ (chosen as large as permitted by avoiding boundary reflections).
The empirical distribution is
\begin{equation}
\widehat{P}(n_0,t)=\frac{1}{R}\sum_{r=1}^{R}\delta_{n_0,n_0^{(r)}(t)},
\label{eq:P_empirical}
\end{equation}
where $\delta_{i,j}$ is the Kronecker delta.
For visualization from a finite ensemble, we further smooth $\widehat{P}(n_0,t)$ using a bounded, boundary-corrected Gaussian
kernel density estimator on the finite support $n_0\in(-n_{\mathrm{LB}}(t),\,n_{\mathrm{LB}}(t))$, in which the Gaussian kernel centered at each sample is truncated to the interval and renormalized to remove boundary bias; the bandwidth is
selected via leave-one-out likelihood cross-validation.

% ============================================================  

% ============================================================
% Figures are placed at the end of the manuscript
% ============================================================
\clearpage
\section{Figures}
\vfill
\vfill

\clearpage

\clearpage

% -----------------------------------------------------------
\clearpage
\begin{figure}[p]
    \centering
    \phantomsection 
    \includegraphics[width=0.99\textwidth]{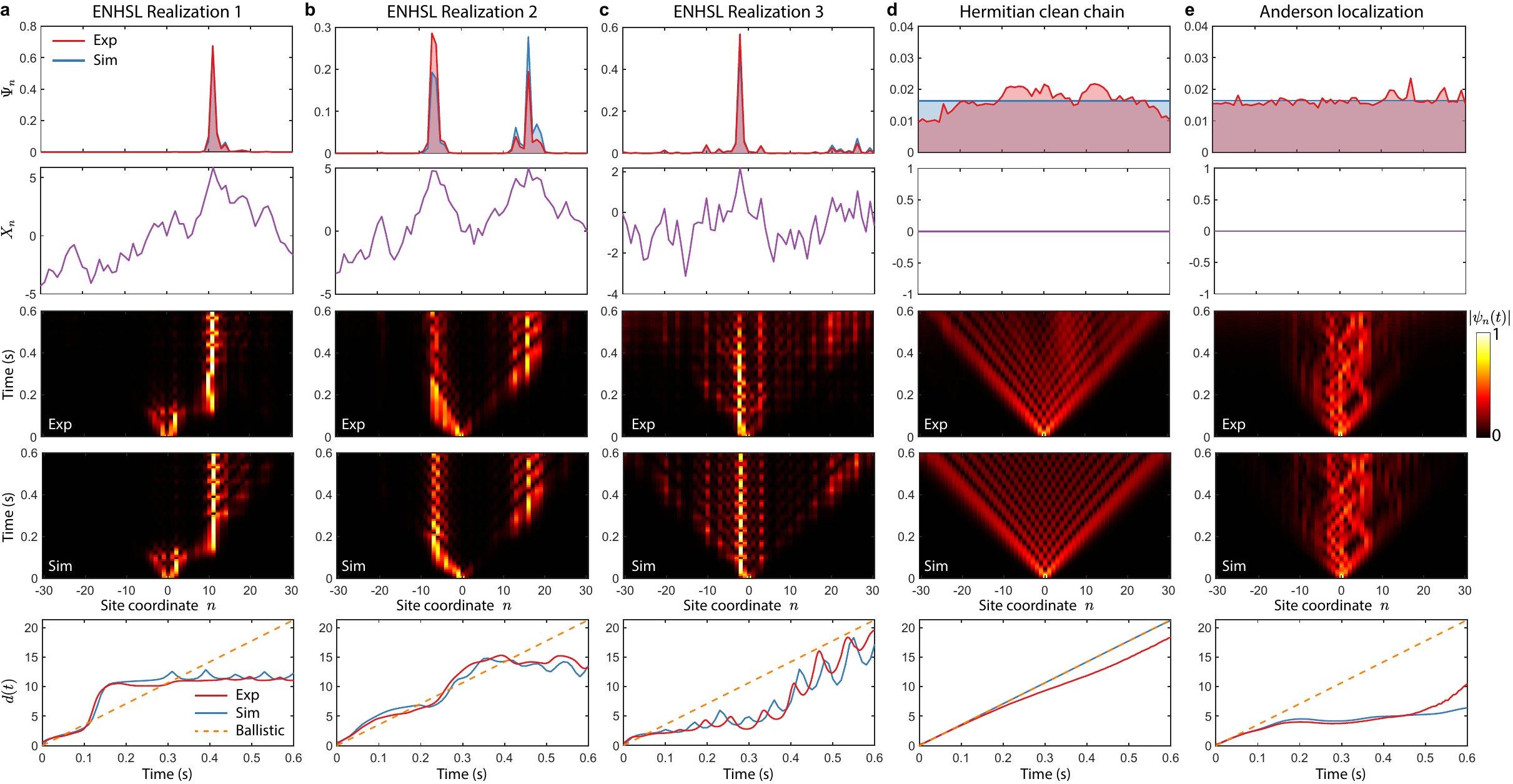}
    \caption{
        \textbf{Experimental observation of wave-packet propagation following excitation at the central site (\(n=0\)) in a chain of length \(N=61\) under OBC.} 
        The parameters set in experiments are \(\omega_0 = 1040\,\mathrm{Hz} - 1.75i\,\mathrm{Hz}\), \(J=4\,\mathrm{Hz}\), and \(\Delta h = 1.6\,\mathrm{Hz}\).
        \textbf{a,b,c}, Three representative disorder realizations of the ENHSL. 
        \textbf{d}, Disorder-free Hermitian chain (clean reference). 
        \textbf{e}, Hermitian chain with onsite disorder showing Anderson localization (reference). For each panel, rows from top to bottom show: (i) the spatial profile of the eigenstate intensity summed over all states, \(\Psi_n=\sum_{m}\lvert\psi_{m,n}\rvert^{2}\); (ii) the random-walk landscape $X_n$; (iii) the experimentally measured spatiotemporal evolution of the acoustic pressure envelope $\abs{\psi_n(t)}$; (iv) the corresponding tight-binding simulation; and (v) the spreading distance \(d(t)\) extracted from experiment (blue solid) and simulation (red solid), with the ballistic law shown for comparison (orange dashed).}
    \label{fig:single_realization_dynamics}
    \addcontentsline{toc}{subsection}{Figure \thefigure}
\end{figure}
%%%%%%%%%%%%%%%%%%%%%%%%%%%%%%%%%%%%%%%%%%%%%%%%%%%%%%%%%%%%%%%%%%%%%%%%%%%%%%%%

\clearpage

% ============================================================

\end{document}